\documentclass[11pt]{article}
\usepackage[a4paper, total={6in, 8in}]{geometry}
\usepackage{lipsum,url}
\usepackage{amssymb}

\usepackage{mathtools}
\usepackage{caption,comment,verbatim}
\usepackage{subcaption}
\usepackage{float}
\usepackage{amsmath}
\usepackage{lastpage}
\usepackage{fancyhdr}

\usepackage{commath}
\usepackage{blindtext}

\title{The Lagrangian formulation for wave motion with a shear current and surface tension }

\author{Conor Curtin $^a$ and Rossen Ivanov $^b$  \thanks{member of the Institute for Advanced Physical Studies, 111 Tsarigradsko shose Blvd., Sofia 1784, Bulgaria}\\
\phantom{*}
\\ School of Mathematics and Statistics, \\Technological University Dublin, \\ City Campus, Grangegorman Lower, \\ Dublin D07 ADY7, Ireland\\
\phantom{8}\\
$^a$ Email: Conor.Curtin@tudublin.ie \\
$^b$ Email: Rossen.Ivanov@tudublin.ie}

\begin{document}

\maketitle

\begin{abstract}
    The Lagrangian formulation for the irrotational wave motion is straightforward and follows from a Lagrangian functional which is the difference between the kinetic and the potential energy of the system. In the case of fluid with constant vorticity, which arises for example when a shear current is present, the separation of the energy into kinetic and potential is not at all obvious and neither is the Lagrangian formulation of the problem. Nevertheless, we use the known Hamiltonian formulation of the problem in this case to obtain the Lagrangian density function, and utilising the Euler-Lagrange equations we proceed to derive some model equations for different propagation regimes. While the long-wave regime reproduces the well known KdV equation, the short- and intermediate long wave regimes lead to highly nonlinear and nonlocal evolution equations.  \\
{\bf Keywords:} Hamiltonian, Dirichlet-Neumann Operator, surface waves, KdV equation, Deep water model.
    \end{abstract}

\section{Introduction}

The aim of this work is to illustrate the possibility of using a Lagrangian instead of Hamiltonian formulation in the derivation of water waves model for a single layer of water in the case of small amplitude for short, intermediate and long waves in the presence of currents. 

Following the seminal work of Zakharov \cite{Zak}, the Hamiltonian approach for water waves propagation has been developed extensively, see for example \cite{Broer,Mil1,Mil2,BO,Rad}. The formulation in \cite{NearlyHamiltonian} shows that the constant vorticity case can be accommodated in the Hamiltonian formulation. A convenient explicit representation of the Hamiltonian involves the non-local Dirichlet-Neumann operator,  e.g. \cite{Craig1993,CraigGroves1,CGK}.  In our derivation we are beginning from the Hamiltonian approach for a single layer gravity waves, with the only assumption for a small wave amplitude compared to the water depth.

The water wave models for a single layer have been in the focus of fluid mechanics from the early days of the scientific research in this area, however the intermediate and short wave models received a lot less attention than the long wave ones, and one reason is perhaps the fact that the corresponding approximations lead to more complicated, nonlinear and nonlocal equations, e.g. \cite{BS71,Mats,CD18,Iv23}.

The short-wave effects usually compete with the capillarity effects and then resonances can be observed - these have been extensively studied, see for example \cite{Phil,MG,Kart,CK,MarJmfm,CaCh,Cra,IM} and will be taken into account in the models under consideration.

In the second Section we describe the setup and the Hamiltonian formulation. In Section 3 we obtain the Lagrangian functional, which produces the equations of motion as Euler-Lagrange equations. The small-amplitude equations are obtained in Section 4 and in Section 5 we proceed and rewrite the equations in evolutionary form. In the shallow water (long-wave) approximation this is a KdV type equation, while in the other regimes the evolution equation is nonlinear and nonlocal. 

\section{Preliminaries}
\subsection{The governing equations}\label{2.1}

\begin{figure}[!ht]
\centering
\includegraphics[width=0.8 \textwidth]{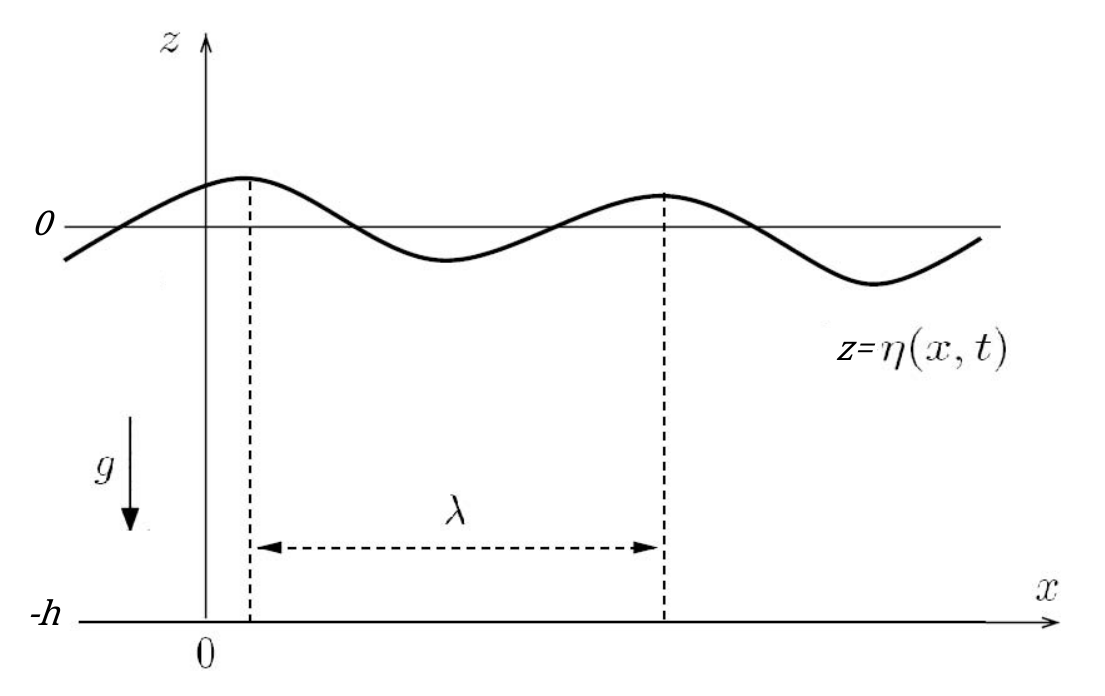}
 \caption{Coordinates and fluid domain.}\label{fig1}
\end{figure}

We choose a Cartesian coordinate system with a horizontal coordinate $x$ and vertical coordinate $z.$
We denote with $t$ the time variable. The governing equations for gravity water waves with a free surface of a two dimensional water flow will be described briefly. The average water surface is at $z=0$ and the wave elevation is given by the function $z=\eta(x,t).$  Therefore we have
\begin{equation}\label{etaAverage}
\int_{\mathbb{R}} \eta(x,t) \, dx =0.
\end{equation}
The fluid domain is bounded below by a flat bed $z=-h$ ($h$ being some positive constant) and above by the surface itself, see Fig. \ref{fig1}.

Denoting with ${\bf V}=(u(x,z,t),0,w(x,z,t))$ the velocity field, which is essentially 2-dimensional; with $P(x,z,t)$ the pressure and with $g$ the gravitational constant, the fluid motion is governed by the Euler equations (we consider for simplicity unit density)
\begin{equation}\label{Eulereq}
\begin{split}
u_t+uu_x+wu_z&=-P_x\\
w_t+uw_x+ww_z&=-P_z-g,
\end{split}
\end{equation}
and the incompressibility equation 
\begin{equation} \label{divV}
  \text{div} {\bf V} = u_x+w_z=0.
\end{equation}
The systems \eqref{Eulereq} and \eqref{divV} are complemented by the boundary conditions
\begin{equation}\label{BKS}
w=\eta_t + u \eta_x\quad{\rm on}\quad z=\eta(x,t),
\end{equation}
\begin{equation}
    w=0\quad{\rm on}\quad z=-h,
\end{equation}
and 
\begin{equation}\label{P}
    P=P_{\rm atm}-\sigma \left( \frac{\eta_x}{\sqrt{1+\eta_x^2}}\right)_x \, \, \text{on} \, \, z=\eta(x,t),
\end{equation} 
were $\sigma$ is the surface tension coefficient (divided by density).

The main assumption is that the velocity field decomposition, according to the Helmholtz Theorem, consists of a potential (irrotational) part and a shear current $U(z)=\gamma z+\kappa$ which is linear in $z$ and brings a constant vorticity $\gamma.$ Indeed, it could be verified that the vorticity is $\nabla \times {\bf V}= \gamma.$  Here $\kappa$ is the current constant value at $z=0.$ Explicitly one can write
\begin{align}\label{uw}
\begin{cases}
u=\varphi_x +U(z) \equiv \varphi_x + \gamma z +\kappa = \psi_z \\
w = \varphi_z=-\psi_x
\end{cases}
\end{align}

Since the fluid is incompressible, from the mass conservation law we have 
\begin{align}\label{masscons}
\nabla \cdot {\bf V} =\left(\varphi_x+\gamma z +\kappa \right)_x + \varphi_{zz}  =\varphi_{xx}+\varphi_{zz}=0. 
\end{align}
Since the wave motion is two-dimensional, from now on we use the notation $\nabla \equiv \nabla^{\perp}=(\partial_x, \partial_z)^T.$  Then the Laplace's equation is satisfied by $\varphi,$ the potential function
\begin{align}
\nabla^2 \varphi = 0 
\end{align}
and also for the case of the streamfunction $\psi$ we have
\begin{align}
\nabla^2 \psi = \psi_{xx}+\psi_{zz} = (-\varphi_z)_x + (\varphi_x +\gamma z +\kappa)_z  = \gamma 
\end{align}
The boundary condition for the bottom of the fluid in terms of potentials may be written as:
$$\varphi_z (x,-h)=0.$$

In terms of the functions $\varphi$ and $\psi$, we can recast the Euler equation \eqref{Eulereq} in the form
$$\nabla \left[ \varphi_t + \frac{1}{2} \,|\nabla \psi|^2 + P - \gamma \psi + g z \right] = 0.$$

Thus \begin{equation}
 \varphi_t + \frac{1}{2} \,|\nabla \psi|^2 + P - \gamma \psi + g z 
\end{equation}  is constant throughout the fluid domain. This is the generalization for flows of constant vorticity of
Bernoulli's law for irrotational flows ($\gamma=0$), see also \cite{NearlyHamiltonian}. In view of \eqref{P}, we deduce that
$$ \varphi_t +
\frac{1}{2} \,|\nabla \psi|^2 -\gamma \psi + g \eta +P_{\rm atm}-\sigma \left( \frac{\eta_x}{\sqrt{1+\eta_x^2}}\right)_x $$ is constant on the
free surface . Since $\varphi$ is uniquely determined by \eqref{uw} up to an arbitrary additive term that is solely time-dependent, we use this freedom to absorb into the definition of $\varphi$ a suitable
time-dependent term so that
\begin{equation}\label{Ber}
\varphi_t+\frac{1}{2} \,|\nabla \psi|^2 -\gamma \psi + g \eta -\sigma \left( \frac{\eta_x}{\sqrt{1+\eta_x^2}}\right)_x= \frac{\kappa^2}{2} .
\end{equation} on the surface, due to the asymptotics of $\psi_z=u\to \kappa$ on the surface. The subindex $s$ will be used as a notation that the evaluation is on the surface.  

Furthermore, we make the assumption that the functions $\eta(x, t),$ and ${\varphi}(x, z, t)$ belong to the Schwartz space of functions $\mathcal{S}(\mathbb{R})$ with respect to the $x$ variable (for any $z$ and $t$). This reflects the localised nature of the wave disturbances. The assumption of course implies that for large absolute values of $x$ the wave attenuates \begin{equation} \label{vanish}
\lim_{|x|\rightarrow \infty}\eta(x,t)=0, \quad \lim_{|x|\rightarrow \infty}{ {\varphi}}(x,z,t)=0 .
\end{equation}
The kinematic boundary condition \eqref{BKS} can be written also as
\begin{align}\label{etaphi}
\eta_t=(\varphi_z)_s-\eta_x\left((\varphi_x)_s +\gamma \eta +\kappa\right).
\end{align}

Next, we introduce the following notation for the evaluation of the potential function at the surface:
$$ \xi(x,t)= \varphi \left(x, \eta(x,t), t \right) = (\varphi)_{s},$$ which belongs to $\mathcal{S}(\mathbb{R})$ with respect to the $x$ variable.

Given the constant vorticity $\gamma$, the functions $\xi$ and $\eta$, completely determine the wave motion.

\subsection{Dirichlet - Neumann Operator, Energy functional, Hamiltonian formulation}
We define now $G(\eta)$, the  Dirichlet - Neumann Operator (DNO) as follows, see for example \cite{CraigGroves1}:
\begin{align} G(\eta)\xi= \left(\frac{\partial \varphi}{\partial {\bf n}} \right)_s \cdot \sqrt{1+\eta_x^2}.
\end{align}
 Here $${\bf n}=\frac{(-\eta_x,1)}{\sqrt{1+\eta_x^2}}$$ is the outward unit normal to the surface,
 \begin{equation}\label{dno1}
 \left(\frac{\partial \varphi}{\partial {\bf n}} \right)_s ={\bf n}\cdot (\nabla  \varphi)_s,   
 \end{equation} 
or following the evaluation of the vector operand
\begin{equation} \label{Gxi}
G(\eta) \xi = -\eta_x (\varphi_x)_s +(\varphi_z)_s.\end{equation}
Then, from \eqref{etaphi}
\begin{align}\label{etat}
\eta_t =G\xi- (\gamma\eta + \kappa)\eta_x .
\end{align}

The energy functional of the system for unit density 
\begin{align*}
E (\xi,\eta) =  \frac{1}{2}\int_{\mathbb{R}} \int_{-h}^{\eta(x,t)} (u^2+w^2)dzdx + g\int_{\mathbb{R}} \int_{-h}^{\eta(x,t)}z\, dzdx +  \sigma \int_{\text{free surface}} dl
\end{align*} can be expressed in terms of $\xi$ and $\eta$ with the help of $G(\eta).$
However, of great importance is the quantity
\begin{align}
\mathcal{H}(\xi, \eta) = E(\xi,\eta) - E(0,0)
\end{align}
which is the relative energy with respect to the unperturbed current state and which plays the role of a Hamiltonian. The calculations follow the routine in \cite{NearlyHamiltonian,CoIv2}. Due to the fact that $|\nabla \varphi|^2=\text{div}(\varphi \nabla \varphi)$ one can use the divergence Theorem for the fluid volume together with the definition of $G(\eta)$ and \eqref{dno1}. The result is
\begin{align*}
\mathcal{H}(\xi, \eta) = & \frac{1}{2} \int_{\mathbb{R}} \xi G(\eta) \xi \, dx - \int_{\mathbb{R}} \xi \eta_x(\gamma \eta+ \kappa)\,dx 
+\frac{1}{6\gamma}\int_{\mathbb{R}}\left[\gamma^3 \eta^3 +3\gamma\kappa \eta^2 \right]\,dx \notag \\
& +\frac{g}{2}\int_{\mathbb{R}}\eta^2\,dx +\sigma \int _{\mathbb{R}} (\sqrt{1+\eta_x^2}-1) dx,
\end{align*}
or, equivalently, expanding the third term and using the fact that due to \eqref{vanish} $$-\int \xi \eta_x(\gamma \eta+\kappa)dx = +\int\left(\gamma \frac{\eta^2}{2}+\kappa\eta\right)\xi_x\,dx,$$ we have another  expression for $\mathcal{H}(\xi, \eta)$ given by:
\begin{align}
\mathcal{H}(\xi, \eta)= &\frac{1}{2} \int_{\mathbb{R}} \xi G(\eta) \xi \, dx + \int_{\mathbb{R}}\left (\frac{\gamma}{2}\eta^2 +\kappa \eta \right) \xi_x \,dx + \frac{\gamma^2}{6}\int_{\mathbb{R}}\eta^3\,dx
\notag \\
& + \frac{g+\kappa \gamma}{2}\int_{\mathbb{R}}\eta^2 dx + \sigma \int _{\mathbb{R}} (\sqrt{1+\eta_x^2} -1)dx. 
\end{align}

The equations of motion, which are equivalent to \eqref{BKS}, \eqref{Ber} have the following "nearly" Hamiltonian form, see \cite{NearlyHamiltonian}:
$$\begin{cases} 
\xi_t = -\frac{\delta \mathcal{H}}{\delta \eta}+\gamma \chi \\
\eta_t= \frac{\delta \mathcal{H}}{\delta \xi}
\end{cases}$$
where  $ \chi\equiv \psi(x,\eta,t) = (\psi)_s .$ 

Our next task will be to write down the system of equations explicitly in terms of  $\xi$ and $\eta$. Important consideration is also given to the consequence and significance of the nearly-canonical Hamiltonian system formulation and as we shall see, it is actually possible to transform it into a canonical Hamiltonian system so that it admits a canonical symplectic structure.

\subsection{The governing equations in terms of $\xi$ and $\eta$ }

We begin by considering our definition of $\chi =(\psi)_s$ and taking it's derivative with respect to $x$:
\begin{align*}
\frac{d \chi}{dx} &=(\psi_x)_s + (\psi_z)_s\eta_x\\
&= -(w-u\eta_x)_s = -\eta_t
\end{align*}
This system admits an integral representation for $\chi, $ given by 
\begin{align}\label{chi00}
\chi = -\int_{-\infty}^{x} \eta_t(x',t)dx' = -\partial_x^{-1}\eta_t.
\end{align} We note that, due to \eqref{etaAverage} 
\begin{equation}
    \partial^{-1}\eta =\int^{x}_{\pm \infty} \eta(x')\,dx'.
    \end{equation}
We note that due to the property \eqref{etaAverage} the result of the integration with both lower limits is the same. However, we now accept the following definition for the negative powers of $D =-i \partial_x$: The action of $D^n$ corresponds to an additional Fourier multiplier $k^n$ in the Fourier representation of the corresponding quantity (for both positive and negative $n$) .
    
Using equation \eqref{chi00} we may write $\chi$ in terms of $\xi$ and $\eta$:
\begin{align}\label{chi}
\chi = -\partial_x^{-1}\eta_t =  -\partial_x^{-1} \left[G(\eta)\xi -\eta_x(\gamma\eta+\kappa)\right]= -\partial_x^{-1} \left[G(\eta)\xi\right] + \gamma \frac{\eta^2}{2}+\kappa \eta.
\end{align}
From the chain rule for $\xi_x(x,t)=\frac{d}{dx}(\varphi(x,\eta,t))$ and \eqref{Gxi} we obtain the system
\begin{equation}\label{dag}
\begin{split}
\xi_x &= (\varphi_x)_s + (\varphi_z)_s \eta_x\\
G\xi &= (\varphi_z)_s - \eta_x (\varphi_x)_s
\end{split}
\end{equation}

Solving system \eqref{dag} simultaneously we obtain the expressions for the gradient components of the surface potential:
\begin{equation}\label{fx}
(\varphi_x)_s = \frac{\xi_x-\eta_xG\xi}{1+\eta_x^2}
\end{equation}
\begin{equation}\label{fz}
(\varphi_z)_s = \frac{G\xi +\eta_x\xi_x}{1+\eta_x^2}
\end{equation}
\par
Our next goal is to use \eqref{fx} and \eqref{fz} to re-formulate Bernoulli's equation \eqref{Ber} and the kinematic boundary conditions \eqref{BKS} in terms of the same variables. 

The evolution equation for $\eta$ is given by \eqref{etat}.

From \eqref{Ber} we have 
\begin{align}
\xi_t -(\varphi_z)_s\eta_t +\frac{1}{2}\left[(\left(\varphi_x)_s +\gamma \eta +\kappa\right)^2 +(\varphi_z)_s^2 \right] -\gamma \chi +g\eta -\sigma \left( \frac{\eta_x}{\sqrt{1+\eta_x^2}}\right)_x = \frac{\kappa^2}{2}.
\end{align}
Then the substitution of \eqref{chi}, \eqref{etat}, \eqref{fx} and \eqref{fz} in the above equation finally gives
\begin{align}
\xi_t + (\gamma\eta+\kappa)\xi_x + \frac{\xi_x^2 -2\eta_x \xi_x G\xi - (G\xi)^2}{2(1+\eta_x^2)} +g\eta -\sigma \left( \frac{\eta_x}{\sqrt{1+\eta_x^2}}\right)_x +\gamma \partial_x^{-1}(G\xi) = 0.
\end{align}

\subsection{Change of Variables and Canonical Hamiltonian formulation}
We introduce a change of variables into the nearly Hamiltonian system given by 
\begin{equation}\label{thetaxi}
\begin{cases}
\theta = \xi + \frac{\gamma}{2} \partial_x^{-1}{\eta} \\
\tilde{\eta}=\eta
\end{cases}
\end{equation}
It is known that in terms of the  new variables the system is Hamiltonian in canonical (standard) form \cite{W} :
\begin{equation}\label{cHam}
\begin{split}
\eta_t &= \frac{\delta \mathcal{H}}{\delta \theta} \\
\theta_t &= -\frac{\delta \mathcal{H}}{\delta \eta}
\end{split}
\end{equation}
where the Hamiltonian is expressed now in terms of $\eta$ and $\theta.$

\section{Lagrangian and Equations of Motion}

We now approach our main goal which is to calculate the Lagrangian of the fluid system, along with the Euler-Lagrange equations of motion. The Lagrangian could be obtained by a Legendre transform of the Hamiltonian \eqref{cHam} and as the medium is continuous the usual summation sign is replaced by integration:
\begin{align}\label{Leg}
\mathcal{L} = \int_{\mathbb{R}} \dot{\eta} \theta \, dx - \mathcal{H}
\end{align}
Here $\theta$ is the generalised momentum of \eqref{cHam}. Since $\cal{L}$ is a function of $\eta $ and $\eta_t,$ we need to express the momentum $\theta$ in terms of the velocity $\dot{\eta}\equiv \eta_t .$ From \eqref{etat} and \eqref{thetaxi} we have a system of two equations:
\begin{align*}
\eta_t &=\dot{\eta} = G\xi -\eta_x(\gamma \eta+\kappa), \\
\xi &= \theta -\frac{\gamma}{2} \partial_x^{-1}\eta.
\end{align*}
Operating on both sides of the first equation with the inverse of the DNO $G^{-1}$ and substituting there $\xi$ from the second equation we obtain
\begin{align*}
G^{-1} \dot{\eta} &= \xi - G^{-1}[\eta_x(\gamma \eta +\kappa)]= \theta -\frac{\gamma}{2}\partial_x^{-1} \eta -G^{-1}[\eta_x(\gamma \eta + \kappa) ]
\end{align*}
and from this the momentum $\theta$ can be expressed as 
\begin{equation}
\theta = \frac{\gamma}{2}\partial_x^{-1}\eta +G^{-1} [\dot{\eta}+\eta_x(\gamma\eta  +\kappa)].
\end{equation}
Furthermore, using the change of variables \eqref{thetaxi} we obtain an expression for $\xi$ as well
\begin{align} \label{xi-eta}
\xi= G^{-1}\left[\dot{\eta} +\eta_x(\gamma \eta + \kappa)\right].
\end{align}
Introducing the following notation reduces the complexity of the expressions:
$$
\Theta = \eta_t + \eta_x(\gamma\eta +\kappa)
$$
this will imply that $\xi$ can be written as 
$$\xi = G^{-1} \Theta.$$
The invertibility of the DNO is discussed rigorously in Appendix A.3 of the book \cite{Lan}.
The DNO is invertible on certain Sobolev spaces. As far as the Schwartz space is a subclass of these spaces, it is sufficient to assume that the physical variables take values in the Schwartz space, see Section 2.1.  \ref{2.1}.

The Hamiltonian functional in terms of our new variables is:
\begin{align}
\tilde{\mathcal{H}}(\eta, \dot{\eta}) &= \frac{1}{2}\int_{\mathbb{R}} \Theta G^{-1} \Theta \,dx +\int_{\mathbb{R}}\left(\frac{\gamma}{2} \eta^2 +\kappa \eta\right)G^{-1} \Theta_x\,dx +\frac{1}{2}(g+\kappa \gamma)\int_{\mathbb{R}}\eta^2 \,dx 
\nonumber \\
&+ \frac{\gamma^2}{6}\int_{\mathbb{R}}\eta^3\,dx +\sigma \int _{\mathbb{R}} (\sqrt{1+\eta_x^2}-1) dx.
\end{align}
The Legendre transform \eqref{Leg} leads to the following explicit expression for the Lagrangian of the system:
\begin{align}
\mathcal{L} (\eta, \dot{\eta}) =& \frac{1}{2} \int_{\mathbb{R}}[\eta_t + \eta_x(\gamma \eta +\kappa)]G^{-1}[\eta_t+\eta_x(\gamma\eta+\kappa)]\,dx \notag \\
&+\frac{\gamma}{2}\int_{\mathbb{R}}\eta_t (\partial_x^{-1}\eta) \,dx -\frac{(g+\gamma\kappa)}{2}\int_{\mathbb{R}}\eta^2\,dx -\frac{\gamma^2}{6}\int_{\mathbb{R}}\eta^3\,dx 
-\sigma \int _{\mathbb{R}} (\sqrt{1+\eta_x^2}-1) dx. \label{Lagrangian}
\end{align}

As one can see the expression is far from intuitive and can not be split in any obvious way into kinetic and potential part. This is apparently due to the complex nature of the fluid motion with vorticity. In the irrotational case of course, the expression \eqref{Lagrangian} reduces to the known formula for the difference between the kinetic and potential energies.

The equations of motion now are reduced to the Euler-Lagrange equation for $\eta:$
\begin{equation}\label{EL}
\frac{d}{dt}\left(\frac{\delta \mathcal{L}}{\delta \eta_t}\right) - \frac{\delta \mathcal{L}}{\delta \eta} = 0.
\end{equation}

The advantage of the method is that this is a single scalar equation for $\eta.$ Then $\xi$ is directly related to $\eta$ due to \eqref{xi-eta}.

In the following sections we derive an explicit evolution equation for $\eta$ in the case of small amplitude approximation.

\section{Water waves over a single layer of fluid with currents - Lagrangian approach}

The result of the previous section will be illustrated by a small amplitude model equation for an arbitrary depth-to-wavelength ratio. To this end we need some details about the Dirichlet-Neumann operators.

\subsection{Expansion of Dirichlet-Neumann Operator}
We introduce the scale parameter, $\varepsilon =a/h$ where $a$ is the wave amplitude ($0<|\eta(x,t)|\le a$).  For small amplitudes or shallow water the scaling is $\varepsilon \ll 1,$ and the Dirichlet-Neumann operator has a perturbative expansion in the form
\begin{align}
G(\eta)= \sum_{j=0}^{\infty}\varepsilon^j G_j (\eta),
\end{align}
as it can be shown that $G(\eta)$ is analytic for $\eta \in \mathbb{C}^{\infty}(\mathbb{R})$ in some ball of radius $O(h)$ around $\eta =0$, \cite{CraigGroves1}.
The first few terms of the expansion of $G(\eta)$ are given by \cite{Craig1993,CraigGroves1}
\begin{align*}
G_0 &=D\tanh(hD),\\
G_1 &=D\eta D -D\tanh(hD)\eta D\tanh(hD),\\
G_2 &=-\frac{1}{2}\left(D^2\eta^2 D\tanh(hD)+D\tanh(hD)\eta^2 D^2 -2D\tanh(hD)\eta D\tanh(hD)\eta D\tanh(hD)\right)
\end{align*}
where $D=-i\partial_x .$
We now may implement a scaling regime based on the recursive form of the DNO and attempt to derive an evolution equation from our Lagrangian formulation. We will need a truncated explicit expression for $G$ and $G^{-1}.$ 

Considering the expansion of $G(\eta)$ up to terms including $\varepsilon^1,$ noting that $\eta$ is of order $\varepsilon,$ (and therefore scales as $\varepsilon \eta$), we write
\begin{align}
G(\eta)=i D \mathcal{T} +\varepsilon \left(D \eta D + D\mathcal{T}\eta D \mathcal{T}\right) + \mathcal{O}(\varepsilon^2),
\end{align}
where, for short we adopt the notation 
\begin{align}
\mathcal{T} :=-i \tanh (hD).
\end{align}
The inverse DNO is therefore  
\begin{align*}
G^{-1}(\eta) = -i D^{-1}\mathcal{T}^{-1} +\varepsilon\left(\mathcal{T}^{-1}\eta \mathcal{T}^{-1} +\eta \right ) + \mathcal{O}(\varepsilon^2).
\end{align*}
Thus using the definition $\partial ^{-1} = -i D^{-1}$ we write 
\begin{align}\label{Ginv}
G^{-1}(\eta) = \partial^{-1} \mathcal{T}^{-1} +\varepsilon \left(\eta + \mathcal{T}^{-1} \eta \mathcal{T}^{-1}\right) + \mathcal{O}(\varepsilon^2).
\end{align}

\subsection{The Lagrangian and the equation of motion}

The contribution of the surface tension term is  
$$\sigma \int _{\mathbb{R}}( \sqrt{1+\eta_x^2} -1)dx  \approx \sigma \int _{\mathbb{R}}\left[\left(1+\frac{1}{2}\eta_x^2 \right) -1\right ] dx =  \frac{\sigma}{2} \int _{\mathbb{R}} \eta_x^2  dx. $$
For simplicity we take $\kappa=0.$ As a matter of fact, the constant $\kappa,$ which is the constant speed of the current on the surface can be removed from the considerations by considering another inertial reference frame, which moves with speed $\kappa,$ that is by considering a change of coordinates $x'=x-\kappa t, $  $t'=t.$  Then expressions like $\eta_t + \kappa \eta_x$ transform to just a time-derivative $\eta_{t'},$ while the $x-$derivatives are unchanged, $\eta_{x'}=\eta_x.$
The corresponding Lagrangian, with the scale parameter written explicitly (assuming all constants like $g,h,\sigma,\gamma$ being of order 1) is

\begin{align}\label{Lagrang2}
\mathcal{L} = &\frac{\varepsilon^2}{2}\int \left(\eta_t  +\varepsilon \gamma \eta \eta_x \right) G^{-1}(\eta)\left(\eta_t +\varepsilon \gamma \eta \eta_x\right)\,dx + \varepsilon^2 \frac{\gamma}{2} \int \eta_t (\partial^{-1}\eta) \, dx  \notag \\
&-\varepsilon^2 \frac{g}{2}\int \eta^2 \, dx -\varepsilon^3 \frac{\gamma^2}{6} \int \eta^3\,dx - \varepsilon^2 \frac{\sigma}{2} \int \eta_x^2 \,dx .
\end{align}

The Euler-Lagrange equation \eqref{EL} with $G^{-1}$ from \eqref{Ginv} produces the equation
\begin{align}\label{Eq1}
\frac{d}{dt} &\left[\mathcal{T}^{-1} \partial^{-1} \eta_t +\frac{\gamma}{2} \partial^{-1} \eta +\varepsilon \left(\frac{\gamma}{2}\mathcal{T} ^{-1}\eta^2 +\eta \eta_t +\mathcal{T}^{-1}(\eta \mathcal{T}^{-1} \eta_t)\right)\right]  \notag \\ 
&- \left[\frac{1}{2} \eta_t^2 -\frac{1}{2}(\mathcal{T}^{-1} \eta_t)^2 -\varepsilon \gamma \eta \mathcal{T}^{-1} \eta_t -\frac{\gamma}{2} (\partial^{-1} \eta_t) -g\eta -\varepsilon \frac{\gamma^2}{2} \eta^2 + \sigma \eta_{xx}\right] = \mathcal{O}(\varepsilon^2). 
\end{align}
There are various identities, involving the operator $\mathcal{T},$ see for example \cite{Sant}. We use the identity
\begin{equation}
    \eta_t \mathcal{T}^{-1} \eta_t=\frac{1}{2} \mathcal{T} (\mathcal{T}^{-1 } \eta_t )^2 - \frac{1}{2} \mathcal{T}(\eta_t^2)
\end{equation}
to simplify \eqref{Eq1} into the form 
\begin{align}\label{Eq2}
\eta_{tt}& +\mathcal{T}\left(\gamma \eta_t +g\eta_x-\sigma\eta_{xxx}\right) -\varepsilon\left[g\eta\eta_x
+g \mathcal{T}(\eta \mathcal{T} \eta_x )-\mathcal{T}(\mathcal{T}^{-1} \eta_t)^2 \right]_x \notag \\
&+\varepsilon\left[\gamma \mathcal{T}(\eta(\mathcal{T}^{-1}-\mathcal{T})\eta_t)+\frac{\gamma^2}{2}\mathcal{T}(\eta^2)\right]_x +\varepsilon \sigma \left[\mathcal{T} \eta \mathcal{T} \eta_{xxx}+\eta \eta_{xxx}\right]_x=\mathcal{O}(\varepsilon^2).
\end{align}
This is a shallow water equation accommodating all possible wavelengths and also taking into account shear current with a linear profile and surface tension. In the special case without current and surface tension ($\gamma=0, \, \sigma=0$) this equation becomes
\begin{align}\label{Eq3}
\eta_{tt}& +g\mathcal{T}\eta_x  -\varepsilon\left[g\eta\eta_x
+g \mathcal{T}(\eta \mathcal{T} \eta_x )-\mathcal{T}(\mathcal{T}^{-1} \eta_t)^2 \right]_x  =\mathcal{O}(\varepsilon^2)
\end{align}
which has been obtained in \cite{Iv23} and is related to the equation obtained by Matsuno in \cite{Mats} in the short-wave regime, when $\mathcal{T}\to \mathcal{H},$ the Hilbert transform.  The equation is second order in $t$, and in addition, it is nonlinear and nonlocal, due to the nature of the operator  $\mathcal{T},$ which is in essence a nonlocal integral function transform. The limit to long- and short wave approximations will be considered in more details in the next section.


\section{Evolutionary form of the model equations}

\subsection{Dispersion Relation}

In what follows we need to explain the action of the nonlocal operators on the functions representing physical quantities. In this case, introducing a Fourier transform in the $x-$variable  ($t$ is not affected by the transform)
\begin{equation}\label{FT}
    \eta(x)=\frac{1}{2\pi} \int e^{ikx} E(k) dk, \qquad E(k)=\int e^{-ikx} \eta(x) dx,
\end{equation}
the action of the operator $\mathcal{A}(D)$ is defined via the corresponding Fourier multiplier as
\begin{equation}
    \mathcal{A}(D) \eta (x) = \frac{1}{2\pi} \int  e^{ikx} \mathcal{A}(k) E(k) dk.
\end{equation}

The dispersion relation of \eqref{Eq2} could be obtained by considering plane wave solutions $\eta(x,t) =\eta_0 e^{i(kx-\omega (k) t)}$ of the linear part of the equation. It satisfies a quadratic equation 
\begin{equation}\label{omegaEq}
    \omega^2+i\gamma\mathcal{T}(k) \omega - ik(g+\sigma k^2) \mathcal{T}(k)=0
\end{equation}
whose solution is 
\begin{align*}
\omega(k)=\frac{1}{2}\left[-\gamma \tanh(kh) \pm \sqrt{\gamma^2\tanh^2(kh)+4k \tanh(kh)(g+\sigma k^2)}\right].
\end{align*}
The wave speed is given by $c = \frac{\omega(k)}{k},$ thus the two possible signs are related to the possibility of a left and right running waves,
\begin{align}\label{c(k)}
c(k)=\frac{1}{2}\left[-\frac{\gamma \tanh(kh)}{k} \pm \sqrt{\frac{\gamma^2\tanh^2(kh)}{k^2}+4 \frac{\tanh(kh)}{k}(g+\sigma k^2)}\right].
\end{align}
Note that $c(k)=c(-k)=c(|k|)$. The definition of the corresponding operator is
\begin{align}\label{c-hat}
\hat{c}(D) =  \frac{1}{2}\left[-i\gamma \frac{\mathcal{T}}{D} \pm \sqrt{-\gamma^2 \frac{\mathcal{T}^2}{D^2}+\frac{4i\mathcal{T}}{D}(g+\sigma D^2)}\right].
\end{align}
In what follows we need also the operator $\hat{c} \partial$  
\begin{align}
\hat{c} \partial =  \frac{1}{2} \left[\gamma\,  \mathcal{T}  \pm \sqrt{\gamma^2 \mathcal{T}^2 -4iD\mathcal{T}(g+\sigma D^2)} \right].
\end{align}

\subsection{Towards an evolutionary equation}
We observe that the model equation \eqref{Eq2} has the following structure 
\begin{align}\label{Bouss}
\eta_{tt}+\gamma \mathcal{T} \eta_t +\mathcal{T}\left(g \eta_x -\sigma \eta_{xxx}\right) +\varepsilon {F}_x=\mathcal{O}(\varepsilon^2),
\end{align}
where
\begin{align}
    {F}=&-g\eta\eta_x -g \mathcal{T}(\eta \mathcal{T} \eta_x ) +\mathcal{T}(\mathcal{T}^{-1} \eta_t)^2 \notag \\
    &+\gamma \mathcal{T}(\eta(\mathcal{T}^{-1}-\mathcal{T})\eta_t)+\frac{\gamma^2}{2}\mathcal{T}(\eta^2) +\sigma \left[\mathcal{T} \eta \mathcal{T} \eta_{xxx}+\eta \eta_{xxx}\right]
\end{align} is a nonlinear functional of $\eta.$
Our aim is to obtain an equation in an evolutionary form 
\begin{align} \label{EE0}
\eta_t +\hat{c}\eta_x +\varepsilon N = \mathcal{O}(\varepsilon^2),  
\end{align}
where $N$ is a nonlinear functional of $\eta$ and which, for example, in a long-wave limit will produce the well known KdV equation \cite{KdV}. 

Taking the time derivative of \eqref{EE0} and using the fact that  $ N_t = -\hat{c} N_x +\mathcal{O}(\varepsilon),$ we obtain 
\begin{align} \label{EE1}
\eta_{tt} +\hat{c}(\eta_t)_{x} - \varepsilon \hat{c} N_x = \mathcal{O}(\varepsilon^2),
\end{align}
From \eqref{Bouss} and \eqref{EE1} we exclude $\eta_{tt}$ and obtain the equation
\begin{align} \label{EE2}
\eta_t -\mathcal{T}(\hat{c}\partial_x -\gamma \mathcal{T})^{-1}(g\eta_x-\sigma \eta_{xxx}) - \varepsilon (\hat{c} \partial_x -\gamma \mathcal{T})^{-1}\left[{F} +  \hat{c}N\right]_x=\mathcal{O}(\varepsilon^2).
\end{align}
It could be checked that 
\begin{align*}
 -\mathcal{T}(\hat{c}\partial_x -\gamma \mathcal{T})^{-1}(g\eta_x-\sigma \eta_{xxx}) \equiv\hat{c}\eta_x 
\end{align*}
thus from the comparison of the nonlinear terms of \eqref{EE2}  and \eqref{EE0} we have
\begin{equation}
   -  (\hat{c} \partial_x -\gamma \mathcal{T})^{-1}\left[{F} +  \hat{c}N\right]_x  =N,
\end{equation}
and finally
%
\begin{align*}
N = -(2\hat{c} -\gamma \mathcal{T} \partial^{-1})^{-1} {F}
\end{align*}
Thus we obtain the evolution equation 
\begin{align}\label{EE3}
\eta_t +\hat{c}\eta_x - \varepsilon(2\hat{c}-\gamma \mathcal{T} \partial^{-1})^{-1}{F}=\mathcal{O}(\varepsilon^2), 
\end{align}
or with the expression for $F,$ where $\eta_t = - \hat{c}\eta_x+\mathcal{O}(\varepsilon)$
\begin{align} \label{EE4}
\eta_t+& \hat{c}\eta_x -\varepsilon(2\hat{c}-\gamma \mathcal{T} \partial^{-1})^{-1}\left [   
-g\eta\eta_x -g \mathcal{T}(\eta \mathcal{T} \eta_x ) +\mathcal{T}(\mathcal{T}^{-1} \hat{c}\eta_x)^2 \right. \notag \\
    & \left. -\gamma \mathcal{T}(\eta(\mathcal{T}^{-1}-\mathcal{T})\hat{c}\eta_x)+\frac{\gamma^2}{2}\mathcal{T}(\eta^2)
    +\sigma \big(\mathcal{T} (\eta \mathcal{T} \eta_{xxx})+\eta \eta_{xxx}\big) \right] =\mathcal{O}(\varepsilon^2).
\end{align}
This is a universal evolution equation, valid for all possible depths and wavelengths. It is nonlinear and highly nonlocal. We point out that there are two expressions for $\hat{c}$ related to the left- and right-running waves, related to the choice of sign in  \eqref{c-hat}. These two expressions correspond to two different equations of the form \eqref{EE4} for these waves.

In what follows we obtain some limiting cases.  

\subsection{Long-wave limit, the KdV equation}

In the long wave limit, where it is often assumed that $\frac{h}{\lambda} \leq 0.05$ and $\varepsilon $  is of order $(\frac{h}{\lambda})^2.$ This means also that $\varepsilon$ is of order $(hk)^2$  or, in operator form, $(hD)^2.$  Therefore we use the expansions
\begin{align}
\mathcal{T} &= -i \tanh (Dh) \rightarrow -\varepsilon ^{1/2} i\left(hD-\varepsilon\frac{D^3 h^3}{3} \right)+\ldots \notag \\  
&= - \varepsilon ^{1/2} \left(h\partial +\varepsilon \frac{1}{3} \partial^3 h^3\right)+\ldots
\end{align}
and therefore $\mathcal{T}(k)$ may be written as:
\begin{align}\label{T(k)}
\mathcal{T}(k)=-i\varepsilon ^{1/2}\left(kh -\varepsilon\frac{1}{3} k^3 h^3\right)+\ldots
\end{align}
We substitute \eqref{T(k)} in the dispersion relation \eqref{omegaEq} and we look for a solution 
\begin{equation}
    \omega(k)=  \varepsilon ^{1/2}(k \omega_1 + \varepsilon k^3 \omega_3)+\ldots,
\end{equation} where $\omega_1$ and $\omega_3$ are constants. The result is
\begin{align}
    \omega_1\equiv & c_0=\frac{1}{2}\left(   -\gamma h \pm \sqrt{\gamma^2 h^2 +4gh}\right) ,\notag \\
    \omega_3 =& \frac{\gamma h^3 \omega_1 -gh^3+3\sigma h}{3(2\omega_1 + \gamma h)}.
\end{align}
We observe that $\omega_1$ is exactly the propagation speed in leading order, satisfying the equation 
\begin{equation}\label{c0}
    c_0^2+\gamma h c_0 - gh =0.
\end{equation}
Thus, the propagation speed is
\begin{equation}
    c(k)=c_0 + \varepsilon k^2 \omega_3 = c_0 + \varepsilon \frac{h^2(\gamma h c_0 -gh)+3\sigma h}{3(2 c_0 + \gamma h)}   k^2   =c_0+\varepsilon \frac{-c_0^2h^2+3\sigma h}{3(2 c_0 + \gamma h)}   k^2
\end{equation}
The corresponding operator is 
\begin{equation}\label{c4KdV}
    \hat{c}=c_0+ \varepsilon \frac{ c_0^2h^2  -3\sigma h}{3(2 c_0 + \gamma h)}   \partial_x^2 .
\end{equation}
The operator $\mathcal{T}$ appears only in the nonlinear part of \eqref{EE4}, where it contributes only by its leading order, 
$\mathcal{T}=-\varepsilon ^{1/2} h\partial.$ For simplifications we use also \eqref{c0}.

The KdV limit of \eqref{EE4} therefore is 
\begin{align} \label{EEKdV}
\eta_t+c_0 \eta_x + \varepsilon \frac{ c_0^2h^2  -3\sigma h}{3(2 c_0 + \gamma h)}\eta_{xxx}
+\varepsilon \frac{3g +\gamma^2h}{2 c_0 + \gamma h}\eta \eta_x=\mathcal{O}(\varepsilon^2).
\end{align}
In the irrotational case $\gamma=0,$ with no surface tension,  $c_0^2=gh$ and we have the familiar result
\begin{align} \label{EE0KdV}
\eta_t+c_0 \eta_x + \varepsilon \frac{ c_0 h^2 }{6}\eta_{xxx}
+\varepsilon \frac{3g }{2 c_0 }\eta \eta_x=\mathcal{O}(\varepsilon^2).
\end{align}
Here we mention that the KdV regime with vorticity in the upper layer and variable bottom has been studied also in \cite{CIMT}.

\subsection{Short-wave limit}

In the short-wave, or ``deep" water approximation it is usually assumed that the ratio $h/\lambda >0.5.$  In this case 
$|\tanh(kh)|\ge \tanh(\pi)= 0.99627, $   that is, we observe that $|\tanh(kh)|$ approaches 1. Therefore we use the approximation $\mathcal{T}=-i\tanh(hD) \approx -i\text{sgn}(D)\equiv \mathcal{H},$  where $\mathcal{H}$ is the well known Hilbert transform, defined by
\begin{equation} \label{HT}
\mathcal{H}\{f\} (x) := \mathrm{P.V.}\frac{1}{\pi}\int_{-\infty}^{\infty}\frac{f(x')dx'}{x-x'}.
 \end{equation}
It has also the property $\mathcal{H}^{-1}=-\mathcal{H},$ therefore the equation in this case becomes 
\begin{align} \label{EE5}
\eta_t+& \hat{c}\eta_x -\varepsilon(2\hat{c}-\gamma \mathcal{H} \partial^{-1})^{-1}\left [   
-g\eta\eta_x -g \mathcal{H}(\eta \mathcal{H} \eta_x ) +\mathcal{H}(\mathcal{H} \hat{c}\eta_x)^2 \right. \notag \\
    & \left. +2\gamma \mathcal{H}(\eta \mathcal{H}  \hat{c}\eta_x)+\frac{\gamma^2}{2}\mathcal{H}(\eta^2)
    +\sigma \big(\mathcal{H} (\eta \mathcal{H} \eta_{xxx})+\eta \eta_{xxx}\big) \right] =\mathcal{O}(\varepsilon^2).
\end{align} where, using the fact that $i \mathcal{H} D = i (-i \text{sgn}(D)) . D =|D|$ which has a Fourier multiplier $|k|,$
\begin{align}
\hat{c}(D) =  \frac{1}{2|D|}\left[-\gamma \pm \sqrt{\gamma^2 +4|D|(g+\sigma D^2)}\right].
\end{align}
In the case when $\gamma=0,$ $\sigma=0$ we have the simplified equation, obtained in \cite{Iv23}
\begin{equation}\label{EE6}
     \eta_t + \hat{c}\eta_x+ \frac{\varepsilon}{2}\hat{c}^{-1}\left[ g \eta \eta_x + g \mathcal{H}(\eta \mathcal{H}\eta_x) -\mathcal{H}(\mathcal{H} \hat{c}\eta_x)^2 \right]=\mathcal{O}(\varepsilon^2), \qquad \hat{c}=\pm \sqrt{g}|D|^{-1/2}.
\end{equation}

These equations are (most likely) not integrable, unlike the integrable KdV equation \eqref{EEKdV}, \eqref{EE0KdV}, they look rather complicated - as a matter of fact these are strictly speaking an integro-differential equation. The possible numeric solution would require a Fourier transform at every time step, then calculation of the corresponding Fourier multipliers and eventually evaluation of all values of the function at the next time step. 

\subsection{Fourier transforms and integral equations}

Considering the Fourier transform \eqref{FT} of $\eta$ in \eqref{EE4} leads to an integro-differential equation for $E(k,t):$
\begin{equation}\label{eqE}
    -iE_t+ k c(k) E(k) +\frac{\varepsilon}{2\pi\left( 2c(k)+\gamma \frac{\tanh(h k)}{k} \right)}\int R(k,q) E(k-q)E(q)\, dq=\mathcal{O}(\varepsilon^2),
\end{equation}
where $c(k)$ is one of the roots \eqref{c(k)} and the kernel $R(k,q)$ is given by
\begin{align}\label{R}
    R(k,q)&=q(g+\sigma q^2) \left[ 1- \tanh(hk) \tanh(hq)\right] \notag \\
    &+\tanh(kh) \left[ \frac{q(k-q) c(q) c(k-q)}{ \tanh(hq)  \tanh(h(k-q))} +\gamma \left(\tanh(hq)+ \frac{1}{\tanh(hq)} \right)q c(q) + \frac{\gamma^2}{2} \right].
\end{align}
This form of the equation may have advantages for the application of numerical methods of solutions.
For the equation \eqref{EE6} the corresponding limit of \eqref{R} is (see also \cite{Iv23})
\begin{equation}
   R(k,q) \to g[q+\text{sgn}(k)(\sqrt{|q||k-q|}-|q|)].
\end{equation}
The numerical solutions of \eqref{eqE} require a suitable discretization of the form (see for example \cite{Rafa})
\begin{equation}
    E(k,t)=\sum_{j}E_j(t)\delta(k-k_j),
\end{equation} for $|k_j| \le k_{\text{max}}$ such that $E(K,t)\approx 0$ for $|k_j| > k_{\text{max}}.$ Then obviously the quantities $E_j(t)$ satisfy a system of ODEs, which is easy to obtain from \eqref{eqE}.  

\subsection{Example }

The model equation \eqref{EE6} is  apparently nonlocal, hence, is not possible to formulate a proper initial value problem. What is possible is to apply a perturbative scheme to generate a solution, starting from some solution 
$\eta^{(0)}(x,t)$ of the linearised equation. Let us take for simplicity 
$$\eta^{(0)}(x,t)=\eta_0 \cos[k_0 (x - c(k) t)]$$ which solves the equation in its leading order (linear approximation), $\eta_0,k_0$ are constants, $k_0$ is the carrier wave-number. Then 
$$E^{(0)}(k,t)=\pi\eta_0(\delta(k-k_0)+\delta(k+k_0))e^{-i k c(k)t}.$$
Looking for a solution 
\begin{align*}
& \eta(x,t)=\eta^{(0)}(x,t)+\varepsilon \eta^{(1)}(x,t)+\ldots, \\
& E(k,t)=E^{(0)}(k,t)+\varepsilon E^{(1)}(k,t)+\ldots
\end{align*}
and noticing that the kernel $R(k,q)$ from \eqref{R} has the property $R(-k,-q)=-R(k,q)$ we employ the Ansatz  
$E^{(1)}(k,t)=E_1(t)(\delta(k-2k_0)+\delta(k+2k_0))$ to obtain a simple ODE for $E_1(t),$ 
\begin{equation*}
    -iE_{1,t}+2k_0 c(2k_0) E_1+ \frac{\pi k_0 \eta_0^2 R(2k_0,k_0)}{2k_0c(2k_0)+\gamma \tanh(2k_0 h)} e^{-2ik_0 c(k_0)t}=0
\end{equation*}
and then the solution
\begin{align}
& E^{(1)}(k,t)=\frac{\pi \eta_0^2 R(2k_0,k_0)}{2 [4k_0 c(2k_0)+\gamma \tanh (2 k_0 h)][c(k_0)-c(2k_0)]}\left(  \delta(k-2k_0)+\delta(k+2k_0)\right)e^{-i k c(k_0)t}, \notag\\
& \eta^{(1)}(x,t)=\frac{ \eta_0^2 R(2k_0,k_0)}{2 [4k_0 c(2k_0)+\gamma \tanh (2 k_0 h)][c(k_0)-c(2k_0)]} \cos[2 k_0 (x - c(k_0)t)]. \notag
\end{align}
This is the second harmonic, and its amplitude is smaller by a factor of $\varepsilon.$ The limit to zero vorticity and surface tension gives the results from \cite{Iv23}. The nonlinearities generate (in principle) all other multiple modes, but the smallness of their amplitudes makes them insignificant.

\section{Conclusions and Discussion}

The results obtained by the Lagrangian approach could be obtained alternatively by using the Hamiltonian approach - see for example the derivation in the irrotational case \cite{Iv23}. The advantage of the Lagrangian approach is that the Euler-Lagrange equation in its standard form gives immediately a scalar equation for the variable $\eta(x,t)$ alone. The minor disadvantage is the relatively complicated, non-intuitive form of the Lagrangian function.
It is of course possible to extend the Lagrangian method to the two-layers case. Results for interfacial internal waves for example are already obtained by the Hamiltonian method, which allegedly has the same range of applicability, see for example \cite{CGK,CIM-16,CoIv2,CuIv} etc. 

From the variable $\eta$ one can reconstruct the $\xi$ variable from \eqref{xi-eta},
and then from the surface variables $(\eta,  \xi)$ the variables like the pressure and the velocity potential $\varphi(x,y)$ in the bulk of the fluid could be recovered.

The nonlinearities could lead in principle to soliton-like solutions. The solitary waves are stable travelling wave solutions which are characterised by a constant speed and fast decay of the wave profile. The solitary waves have a different nature in comparison to the periodic waves. 
For the integrable KdV approximation (for waves on shallow water) these are its famous solitons, see for example the inverse-scattering method for the KdV equation in \cite{ZMNP}.
The existence of solitary waves usually depends on the dispersion relation - and the possibility for a real propagation speed for imaginary wave-number $k.$
In this regard, the presence of surface tension and current leads indeed to the existence of solitary waves on deep water. This has been proven in \cite{Ioos}, although numerically these have been studied previously for example in \cite{LH,VdB}.

\subsection*{Conflict of interest statement}

On behalf of all authors, the corresponding author states that there is no conflict of interest.

\subsection*{Data availability statement }

The reported results are of purely theoretical nature and no data has been used supporting these results.

\subsection*{Acknowledgements} The authors are thankful to two anonymous referees for their corrections and constructive suggestions, which have improved the quality of the manuscript. This publication has emanated from research conducted with the financial support of Science Foundation Ireland under Grant number 21/FFP-A/9150. 


\end{document}